\documentclass{iau_FM}
\usepackage{graphicx}

\title[Intergalactic Magnetic Field with the CTA] %% give here short title %%
{Large-Scale Diffuse Intergalactic Magnetic Fields Constraints with the Cherenkov Telescope Array} 

\author[P. Barai \& E. M. de Gouveia Dal Pino (CTA Collaboration)]   %% give here short author list %% 
{Paramita Barai$^1$, Elisabete M. de Gouveia Dal Pino$^1$ (on behalf of the CTA Collaboration)} 

\affiliation{ 
$^1$Instituto de Astronomia, Geof\'isica e Ci\^encias Atmosf\'ericas - 
Universidade de S\~ao Paulo (IAG-USP), 
Rua do Mat\~ao 1226, S\~ao Paulo, 05508-090, Brazil \\ email: {\tt paramita.barai@iag.usp.br} 
} 

\pubyear{2018} 
\jname{Astronomy in Focus} 
\editors{Teresa Lago, ed.} 

\begin{document} 

\maketitle 

\begin{abstract} 
Magnetic fields of the order of $\mu$-Gauss are observationally detected 
in galaxies and galaxy clusters, which can be (at least) in part originated by 
the amplification of much weaker primordial seed fields. 
These fields should be carried out by strong galactic outflows, 
magnetically enriching the InterGalactic Medium (IGM). 
However direct observation of magnetic fields in the IGM is scarce. 
This talk will give a review of how Intergalactic Magnetic Field (IGMF) 
can be constrained using gamma-ray observations. 
High-energy TeV photons emitted by distant blazars can interact with the 
cosmic extragalactic optical/infrared/microwave background light, producing 
electron-positron pairs, and initiating electromagnetic cascades in the IGM. 
The charged component of these cascades is deflected by IGMFs, 
thereby reducing the observed point-like TeV flux, and creating an extended image 
in the GeV energy range, which can potentially be detected with $\gamma$-ray telescopes 
(Fermi-LAT, HESS, CTA). Studies (e.g., \cite{Neronov10, Dolag11}) have put lower limits 
on the IGMF strength of the order of $10^{-16} - 10^{-15} G$, and filling factors of $60\%$. 
This talk will describe the constraints which the Cherenkov Telescope Array 
sensitivity is expected to give (\cite{CTA18}). 
\end{abstract} 

\firstsection % if your document starts with a section,

\section{Introduction} 
\label{sec-Introduction} 

Magnetic Fields (MF) of the order of $(1 - 10) \mu G$ are observed 
(e.g., \cite{Berkhuijsen03}) in galaxies and galaxy clusters. 
These fields are understood to be created from the amplification of 
much weaker primordial seed fields, by turbulent dynamo effects and baryonic processes. 
The origin of these tiny seed fields is unknown (e.g., \cite{Widrow02}). 
There are 2 classes of concordance models for the generation of primordial seed fields. 
One is {\it Cosmological origin}: where the seed fields are produced 
in the early primordial Universe (e.g., \cite{Grasso01}). 
The other is {\it Astrophysical origin}: where the seed fields are produced 
by plasma motions from baryonic processes 
(star-formation, supernovae, black holes) in galaxies (e.g., \cite{Ryu12}). 

% during the epochs of inflation or decoupling, by some kind of exotic process like phase transitions 

%%%%%%%%%%%%%%%%%%%%%%%%%%%%%%%%%%%%%%%%%%%%%%%%%%%%%%%%%%%%%%%%%%%%%%%% 
% FIGURE 1 
\begin{figure} 
\centering 
\includegraphics[width = 0.9 \linewidth]{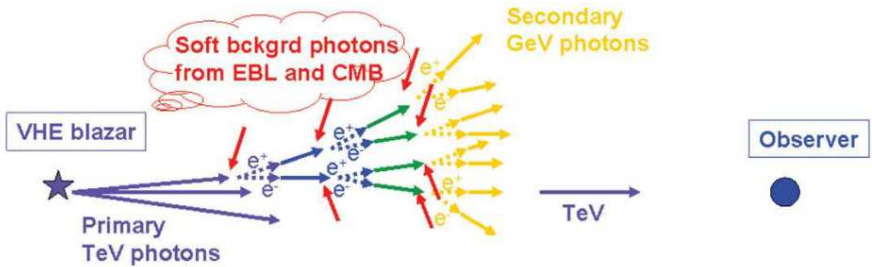} 
\caption{ 
Propagation of VHE primary TeV photons $\gamma_{\rm VHE,1}$ to Earth, 
and the interactions happening on the way to produce 
secondary GeV $\gamma_{\rm VHE,2}$ photons. Figure from Sol et al. (2013a). 
} 
\label{Fig1_Sol_2013} 
\end{figure} 
%%%%%%%%%%%%%%%%%%%%%%%%%%%%%%%%%%%%%%%%%%%%%%%%%%%%%%%%%%%%%%%%%%%%%%%% 

A potentially relevant component of cosmic MF, of which very little is known yet, 
is the Intergalactic Magnetic Field (IGMF) existing in the 
low-density InterGalactic Medium (IGM), 
or the space between galaxies not related to gravitational collapse. 
The MFs inside galaxies are dragged out by powerful galactic outflows 
generated by energy feedback from baryonic processes with time, 
and dispersed into the IGM. 
Over cosmological epochs these IGMFs permeate the turbulent IGM, 
and may be even amplified by turbulent dynamo action (e.g., \cite{deGouveia11}), 
at the same time influencing large scale structure formation (e.g., \cite{Dolag06, Barai08}). 

% IGMF is coherent on scales larger than structures (galaxies and clusters) in the cosmos. 
% ; together with large-scale 
% dynamical events, e.g.~galaxy mergers, and tidal interactions between galaxies. 

The knowledge of IGMF distribution is crucial in understanding the 
{\it cosmological} versus {\it astrophysical} origin of cosmic MF. 
It is challenging to directly observe the IGMF, 
because they are diffuse and weak in intensity. 
An upper limit: $B_{IGM} < (10^{-8} - 10^{-9}) G$, 
is provided by standard constraints: Big Bang nucleosynthesis, 
Cosmic Microwave Background (CMB) anisotropy (e.g., \cite{Durrer98}), 
Faraday rotation measures of polarized radio emission from quasars 
(e.g., \cite{Pshirkov16}). 
Here we will overview how IGMF can be constrained using $\gamma$-ray observations, 
which provide lower limits for it. 

% stronger primordial MF would change the expansion rate of our Universe, 
% and hence the abundance of the primordial elements. 
% MF present at the decoupling epoch induces unequal expansion in 
% different directions and would distort the 
% which is not observed over $10$Mpc-scale 

\section{IGMF Constraints using $\gamma$-ray Observations of Blazars} 
\label{sec-IGMF-using-Blazars} 

A novel technique to constrain the IGMF's strength and filling factor uses 
Very-High-Energy (VHE) $\gamma$-ray emission from distant blazar sources. 
Blazars are Active Galactic Nuclei (AGN) with the central supermassive black hole jet 
pointed toward our line of sight, and present rapid variability (e.g., \cite{Aharonian07}). 
The observed spectral energy distributions of blazars (e.g., \cite{Bonnoli15}) 
are fitted to models, where the TeV $\gamma$-ray emission comes from the jet base. 
There, relativistic electrons ($e^{-}$) upscatter, by Inverse Compton (IC), 
lower-energy ambient photons to the TeV. 
Relativistic protons can also have a contruibution, by emitting direct synchrotron radiation, 
or by the creation of secondary pions, which decay into TeV photons. 

The VHE primary TeV photons ($\gamma_{\rm VHE,1}$) emitted by distant blazars 
undergo the following interactions, as they travel 100s-of-Mpc 
intergalactic space before reaching Earth: 
\begin{equation} 
\gamma_{\rm VHE,1} + \gamma_{\rm EBL} \longrightarrow e_{\rm 1}^{-} e_{\rm 1}^{+}, 
~~~~~~~~~~~~~ 
e_{\rm 1}^{-} + \gamma_{\rm CMB} \longrightarrow \gamma_{\rm VHE,2} + e_{\rm 2}^{-} ~. 
\end{equation} 
Firstly, the $\gamma_{\rm VHE,1}$ interact with optical/infrared/microwave 
Extragalactic Background Light (EBL), and cause $e^{-} e^{+}$ pair production. 
These very-energetic electrons undergo IC scattering off CMB photons 
and produce secondary $\gamma$-rays ($\gamma_{\rm VHE,2}$) in the GeV energy range, 
as shown in Fig.~\ref{Fig1_Sol_2013} (\cite{Sol13a}). 
The $e^{-} e^{+}$ pair electromagnetic cascades are deflected by IGMF present 
in the intervening space, and the secondary GeV $\gamma_{\rm VHE,2}$ 
are strongly attenuated as they appear on Earth. 
These attenuated secondary GeV components can be detected with 
our $\gamma$-ray telescopes (Fermi-LAT, HESS, CTA), as either: 
\begin{itemize} 
\item {\bf Pair Halo}: 
Spatially-extended GeV emission around primary TeV $\gamma_{\rm VHE,1}$ signal. 
These are expected for $B_{IGM} >\sim 10^{-16} G$, and involve 
imaging analysis searches for extended pair halos around blazars. 
(Larger IGMFs produce larger deflections, resulting in a weaker pair halo flux, 
that can make it undetectable with current instruments.) 
\item {\bf Pair Echo}: 
GeV emission with a time delay relative to the primary. 
These are expected for $B_{IGM} < 10^{-16} G$, and involve 
time-resolved spectral analysis of pair echoes. 
\end{itemize} 

Studies usually model the $e^{-} e^{+}$ pair cascade development 
using Monte-Carlo simulations, compute the simulated {\bf pair halo} 
and/or {\bf pair echo} assuming some IGMF configuration, 
compare with observations (e.g. Fermi data on blazars), 
and derive IGMF constraints (e.g., \cite{AlvesBatista17}). 
However, several studies have inferred a non-detection of secondary components, 
which nevertheless provide lower limits on $B_{IGM}$, assuming that the 
suppression of GeV flux is due to the deflection of $e^{-} e^{+}$ pairs by IGMF. 
E.g. Neronov \& Vovk (2010) found a lower limit of $B_{IGM} \geq 3 \times 10^{-16} G$. 
Dolag et al. (2011) inferred that IGMF fills at least $60\%$ of space 
with fields stronger than $B_{IGM} \geq 10^{-16} - 10^{-15} G$. 
Considering a coherence length $> 1$ Mpc for the IGMF and persistent TeV 
emission, \cite{Taylor11} inferred a $B_{IGM} > (10^{-17} - 10^{-15}) G$. 

% if dimming of the cascade emission is due to spatial extension, 
% and $B_{IGM} > 10^{-17} G$ if it is due to time delay. 

A first hint for the existence of pair halos (extended emission around a 
point-source) has been found by \cite{Chen15}, by the stacking analysis of 
$24$ blazars at $z < 0.5$ using Fermi-LAT data. 
It implies a magnetic field strength of $B_{IGM} \sim 10^{-17} - 10^{-15} G$, 
using a Bayesian statistics.

\section{$\gamma$-ray Pair Halo to be Observed by the Cherenkov Telescope Array} 
\label{sec-Pair-Halo-CTA} 

The Cherenkov Telescope Array (CTA) is a planned next-generation ground-based 
$\gamma$-ray observatory (\cite{CTA18}). 
CTA will provide the deepest insight into the non-thermal VHE Universe 
ever reached, probing the physical conditions of 
cosmic accelerators like black holes and supernovae. 
It will consist of an array of around $100$ 
imaging atmospheric Cherenkov telescopes of various sizes. 
CTA foresees a factor of $5 - 10$ improvement in sensitivity in the energy domain 
($\sim 100$ GeV $- 10$ TeV) of current Cherenkov telescopes: HESS, MAGIC, and VERITAS. 
It is also expected to extend the observable VHE range 
to below $100$ GeV and above $100$ TeV, and have unprecedented angular 
and energy resolution, as well as a wider field-of-view. 
A Northern site (Canarias Islands) and a Sourthern site (Chile) 
are planned for a full sky coverage. 

\begin{figure} 
\centering 
\includegraphics[width = 0.47 \linewidth]{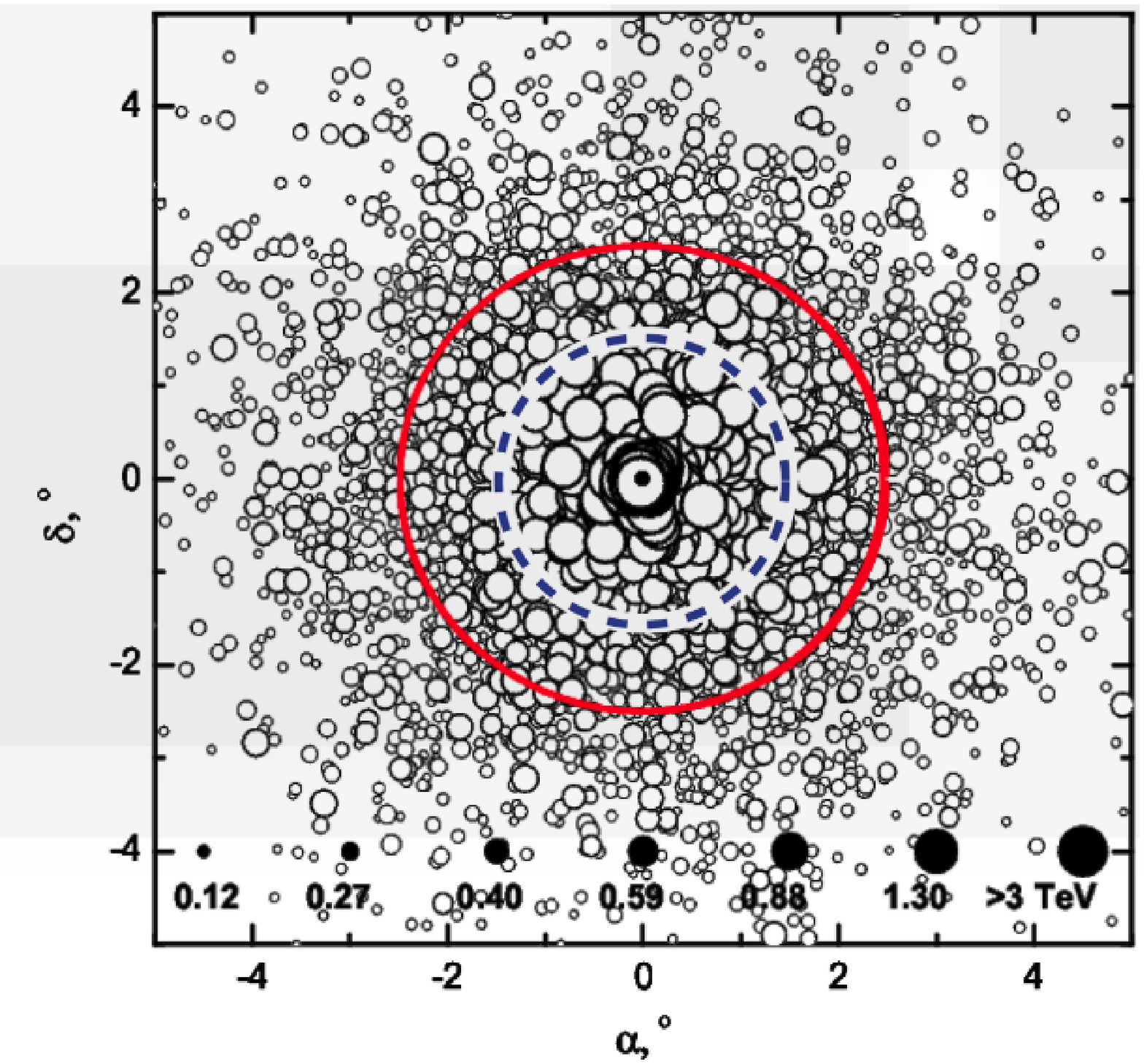} 
\hspace{0.1cm} 
\includegraphics[width = 0.50 \linewidth]{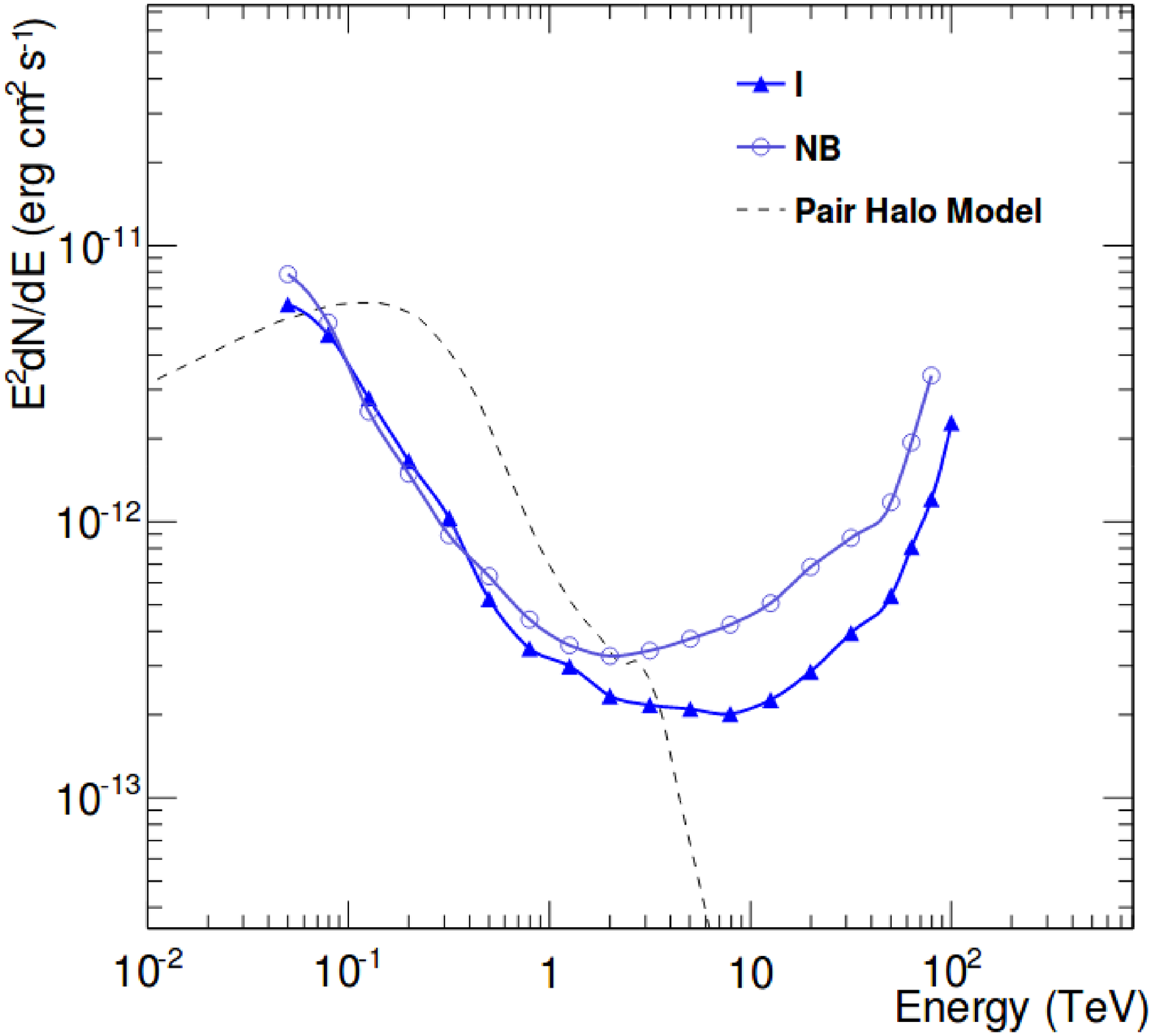} 
\caption{ 
{\it Left panel}: 
Open black circles indicate the arrival directions of primary and secondary 
$\gamma$-rays from a blazar (details in \S\ref{sec-Pair-Halo-CTA}), 
with the circle sizes proportional to the photon energies 
(as labeled in TeV at the bottom of the figure). 
Sky FoVs of $1.5^{\circ}$ and $2.5^{\circ}$ are shown by the 
blue-dashed and red-solid circles. Figure from \cite{Elyiv09}. 
{\it Right panel}: 
Estimated flux for the expected pair halo emission compared to the CTA 
sensitivity curves for the southern (I) and northern (NB) sites. 
Figure from CTA Consortium (2018), originally from \cite{Sol13b}. 
} 
\label{Fig2_Fig3} 
\end{figure}

\cite{Elyiv09} performed Monte Carlo simulations of 3D electromagnetic cascades 
(described in \S\ref{sec-IGMF-using-Blazars}). 
One of the cases they simulated is a blazar source at a distance $120$ Mpc, 
and $B_{IGM} = 10^{-14} G$. 
The expected geometry of {\it Pair Halo} from this blazar is presented 
in Fig.~\ref{Fig2_Fig3} - left panel. 
The sky Field-of-View (FoV) of $1.5^{\circ}$ is indicated by the blue-dashed circle, 
which is equal to the radius of the FoV of the MAGIC telescope. 
And a $2.5^{\circ}$ FoV is denoted by the red-solid circle, 
which corresponds to the size of the FoV of the HESS telescope. 
These would perfectly fit into the much larger FoV of CTA. 

% whose intrinsic $\gamma$-ray spectrum is described as a power law with an exponential cutoff, 

%%%%%%%%%%%%%%%%%%%%%%%%%%%%%%%%%%%%%%%%%%%%%%%%%%%%%%%%%%%%%%%%%%%%%%%% 
% FIGURE 2 & 3 
%%%%%%%%%%%%%%%%%%%%%%%%%%%%%%%%%%%%%%%%%%%%%%%%%%%%%%%%%%%%%%%%%%%%%%%% 

\cite{Sol13b} estimated the pair halo flux 
using a theoretical model of differential angular distribution of a pair halo 
at $z = 0.129$ with $E_{\gamma} > 100$ GeV (\cite{Eungwanichayapant09}), 
and assuming an observation time of $50$ hours. 
The pair halo emission is displayed in Fig.~\ref{Fig2_Fig3} - right panel, 
as the dashed curve. 
The CTA sensitivity curves are overplotted as the solid lines: 
the CTA South (North) site is labeled as I (NB). 
Hence CTA should well observe pair halos in the energy range $(0.1 - 5)$ TeV.

\section{Conclusions} 
\label{sec-Conclusions} 

The observational detection of IGMF (extremely tiny magnetic fields 
permeating the cosmos on the largest scales) 
can shed light on the origin of seed fields in the Universe. 
Current results of VHE $\gamma$-ray astronomy conclude to 
the existence of a non-zero IGMF: $10^{-17} G < B_{IGM} <  10^{-14} G$, 
mostly based on the non-detection of expected secondary GeV $\gamma$-rays. 
Future theoretical studies need to take into account possible additional 
effects in the IGM, e.g. energy losses by cosmic rays, plasma effects. 
Observationally the positive detection of {\it Pair Halos} and {\it Pair Echos} 
are needed with detailed data on cascade signatures, 
which the CTA with its improved sensitivity is expected to observe.

\section{Acknowledgements} 

This work is partially supported by the FAPESP grants 
\#2016/01355-5 and \#2013/10559-5, and by a CNPq grant (\#306598/2009-4).

\end{document}